\documentclass[prb,superscriptaddress,twocolumn]{revtex4-2}
\usepackage[utf8]{inputenc}

\usepackage{hyperref}                                                   % for clickable links inside the pdf - both crossreferences and links to the internet
\hypersetup{
  colorlinks   = true, %Colours links instead of ugly boxes
  urlcolor     = cyan, %Colour for external hyperlinks
  linkcolor    = blue, %Colour of internal links
  citecolor    = blue  %Colour of citations, could be ``red''
}
\usepackage{color}
\usepackage{graphicx}
\usepackage{xcolor,color}
\usepackage{bm,braket}

\usepackage{amsmath}
\usepackage{amsfonts}

\usepackage{mathtools}
\usepackage{tikz}
\tikzset{dot/.style={draw=black,fill=white,circle,inner sep=1pt}}
\DeclareRobustCommand\circle[1]{%
  \thinspace\smash{\tikz[baseline=(A.base)]\node[dot](A){#1};}\thinspace%
}                                 

\begin{document}

\title[Magnetic properties in infinite-layer nickelates]{Magnetic properties and pseudogap formation in infinite-layer nickelates: insights from the single-band Hubbard model} 

\author{Marcel Klett}
\email{m.klett@fkf.mpg.de}
\affiliation{Max-Planck-Institut für Festk{\" o}rperforschung, Heisenbergstraße 1, 70569 Stuttgart, Germany}
\author{Philipp Hansmann}
\affiliation{Department of Physics, Friedrich-Alexander-University (FAU) Erlangen-N{\"u}rnberg, 91058, Erlangen, Germany}
\author{Thomas Sch\"afer}
\affiliation{Max-Planck-Institut für Festk{\" o}rperforschung, Heisenbergstraße 1, 70569 Stuttgart, Germany}

%%%%%%%%%%%%%%%%%%%%%%%%%%%%%%%%%%%%%%%%%%%%%%%%%%%%%%%%%%%%%%%
%%%%%%%%%%%%%%%%%%%%%%%%%%%%%%%%%%%%%%%%%%%%%%%%%%%%%%%%%%%%%%%
\begin{abstract}
We study the magnetic and spectral properties of a single-band Hubbard model for the infinite-layer nickelate compound LaNiO$_2$. As spatial correlations turn out to be the key ingredient for understanding its physics, we use two complementary extensions of the dynamical mean-field theory to take them into account: the cellular dynamical mean-field theory and the dynamical vertex approximation. Additionally to the systematic analysis of the doping dependence of the non-Curie-Weiss behavior of the uniform magnetic susceptibility, we provide insight into its relation to the formation of a pseudogap regime by the calculation of the one-particle spectral function and the magnetic correlation length. The latter is of the order of a few lattice spacings when the pseudogap opens, indicating a strong-coupling pseudogap formation in analogy to cuprates.
\end{abstract}

\maketitle

%%%%%%%%%%%%%%%%%%%%%%%%%%%%%%%%%%%%%%%%%%%%%%%%%%%%%%%%%%%%%%%
%%%%%%%%%%%%%%%%%%%%%%%%%%%%%%%%%%%%%%%%%%%%%%%%%%%%%%%%%%%%%%%
\section{Introduction}
\label{sec:intro}

With the discovery of superconductivity in Sr-doped NdNiO$_2$ in 2018 \citep{li19} it is likely that a new branch of the family of unconventional superconductors (i.e. with non-phonon mediated pairing) was revealed. At this time nickelates, as bulk materials and heterostructures, have already been in the focus of an intense search for high-$T_\text{c}$ cuprate anaolgue oxides for a while (see e.g. \citep{anisimov99, lee04,giniyat08,hansmann09, hansmann10, benckiser11, han11,disa15}). 
One of the current challenges is therefore to understand similarities and/or differences between nickelate and other unconventional superconductors like, e.g., cuprate-, organic-, iron pnictide-, and heavy-fermion compounds. 
While there is currently no consensus if these materials could be covered by a single theory, there are strong indications that for all of them purely electronic (in particular magnetic) fluctuations are at least part of the key to understand their pairing mechanism.
Such fluctuations are also expected to be responsible for unusual observations above the critical temperature which for the high-$T_\text{c}$ cuprates include non-Fermi liquid behaviour in i) temperature dependence of resistivity (universal in all cuprates, e.g., \cite{Takagi1992,Taillefer09}, and found also in organic- and iron pnictide-SC \cite{doiron09,Taillefer10} ii) magnetic susceptibilities which are neither Pauli- nor Curie-like but exhibit sharp drops at a new temperature scale commonly denoted $T^*$ \cite{Alloul89}, and iii) partially (i.e. momentum dependently) gapped quasi-particle Fermi surfaces \cite{Shen2005,Kanigel2006,Damascelli2003}.
The region of these phenomena in the temperature/hole-doping phase diagram is commonly referred to as the ``pseudogap" region.

Motivated by our recent combined experiment/theory multi-method study of the static uniform magnetic susceptibility $\chi$ in LaNiO$_2$ \cite{Ortiz2021} and other recent experimental studies \cite{Zhao2021}, in this manuscript we investigate deeper how the two-particle magnetic response is linked to one-particle spectra $A(\mathbf{k},\omega)$ for different temperatures and different doping levels. 
With the help of complementary quantum many-body techniques, we show that the emergence of a maximum in $\chi$ is concomitant with a significant drop in the antinodal weight of $A(\mathbf{k}\!\approx\!(\pi,0),\omega\!=\!\varepsilon_F)$ at the Fermi level. On the basis of these results we argue that - like cuprates - also nickelate superconductors feature a pseudogap region in their phase diagram.

The paper is organized as follows: in Sec.~\ref{sec:model} we introduce the effective single-band model of infinite-layer nickelates and a brief overview of the numerical methods used to analyze it. In Sec.~\ref{sec:results} we present our results starting with the temperature/doping phase diagram obtained from the maxima of $\chi$ (Sec.~\ref{sec:phase_diagram}). Afterwards we show the one-particle spectral functions (Sec.~\ref{sec:spectral}) and provide magnetic correlation lengths as a function of temperature (Sec.~\ref{sec:xi}). We conclude the paper in Sec.~\ref{sec:conclusions} by commenting on the relevance of our findings to infinite-layer nickelates and their cuprate analogues.

\begin{figure*}[t!]
\centering
\includegraphics[width=0.85\textwidth]{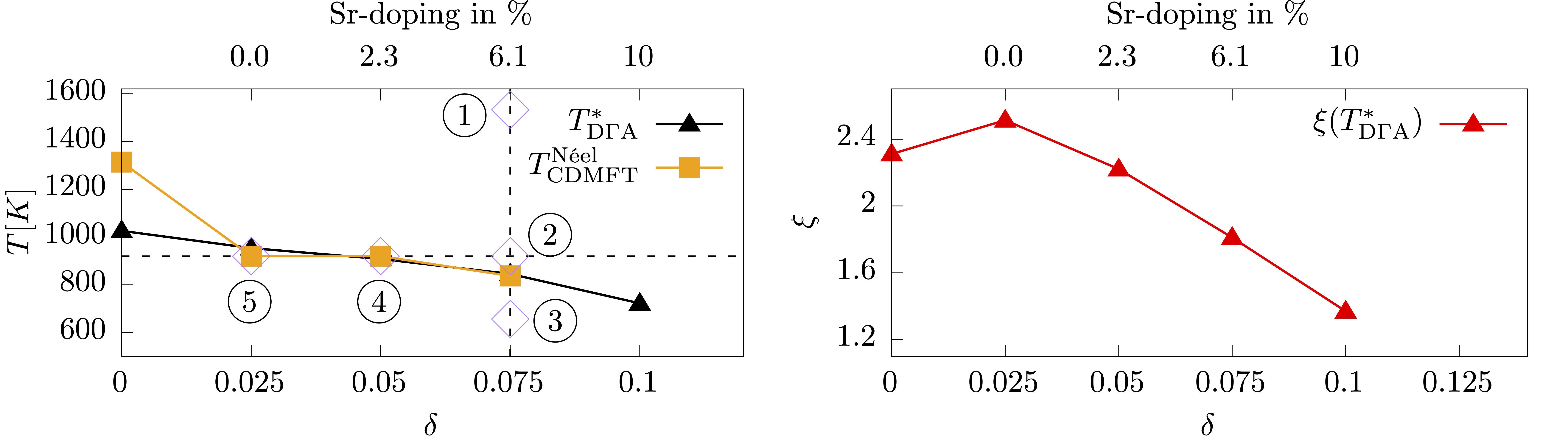}
\caption{Left: Phase diagram of the Hubbard model given by Eq.~(\ref{eq:Hubbard}) as a function of temperature $T$ and doping $\delta$. $T^\ast$ indicates the maximum of $\chi$ calculated in  D$\Gamma$A (black triangles). The orange squares indicate the magnetic ordering temperature in CDMFT $T^{\text{N{\' e}el}}_\text{CDMFT}$, signalling the onset of non-local correlations. The points \circle{1}--\circle{5} (diamonds) refer to Fig.~\ref{fig:3} and  Fig.~\ref{fig:4}. Right: The correlation length $\xi$ in units of lattice spacings is shown for the temperatures $T^*$.\label{fig:1}}
\end{figure*}
\section{Model and methods}
\label{sec:model}
For our study we use the single-band Hubbard model 
\cite{Hubbard1963,Hubbard1964,Kanamori1963,Gutzwiller1963,Qin2022,Arovas2022} on a two-dimensional square lattice:
\begin{equation}
    H =  - \sum_{\langle i,j \rangle} \sum_{\sigma} t_{i,j} \, \hat{c}^\dagger_{i,\sigma} \hat{c}_{j,\sigma} - \mu \sum_{i}  \sum_{\sigma} \hat{n}_{i,\sigma}  +  U \sum_{i}  \hat{n}_{i,\uparrow} n_{i,\downarrow},
    \label{eq:Hubbard}
\end{equation}
where $\sigma$ is the spin of the electron, $\hat{c}^\dagger_{i,\sigma}$ ($\hat{c}_{i,\sigma}$) creates (annihilates) an electron on lattice site $i$ with spin $\sigma$ and $\hat{n}_{i,\sigma}$ is the number operator.

This model has already been successfully applied in the description of the superconducting phase in NdNiO$_2$ \cite{Kitatani2020} and the non-Curie-Weiss behavior of the magnetic susceptibility in LaNiO$_2$ \cite{Ortiz2021}. The material realistic hopping parameters, resulting from a Wannier- and tight-binding projection are $t\!=\!395 \mathrm{meV}$, $t'\!=\!-0.25t\!=\!-95~\mathrm{meV}$ and $t''\!=\!0.12t\!=\!47~ \mathrm{meV}$ as well as the local Hubbard interaction $U\!=\!8t\!=\!3.16 ~\mathrm{eV}$ from a cRPA calculation \cite{Kitatani2020}. All energies are given in units of eV except for temperatures, which are given in Kelvin. The chemical potential $\mu$ is adjusted to an average filling of $n = 1- \delta$, where $\delta$ indicates the hole-doping of the single $d_{x^2-y^2}$ band. When relevant we will also give the corresponding level of Sr-doping.

We investigate the properties of the model in Eq.~(\ref{eq:Hubbard}) as a function of temperature $T$ and doping $\delta$ by applying three numerical methods. Besides dynamical mean-field theory (DMFT) \cite{Georges1996,Georges1992,Metzner1989}, which includes all temporal onsite-correlations of the lattice problem, we use two complementary extensions of it: cellular dynamical mean-field theory (CDMFT) \cite{Maier2005} and the dynamical vertex approximation (D$\Gamma$A) \cite{Toschi2007,Katanin2009}, a diagrammatic extension of DMFT \cite{Rohringer2018}. The combination of complementary numerical methods (`multi-method approach', \cite{Schaefer2021, LeBlanc2015}) turned out to be very useful and versatile recently for both purely model- \cite{Schaefer2021,Wietek2021} and material-based \cite{Ortiz2021} studies. 

For the present work we make use of this approach in order to study the influence of (non-local) magnetic fluctuations captured by the different approximations on different length scales. CDMFT is a conceptually simple real-space cluster extension of DMFT and controlled in the sense that it recovers the exact solution for infinite cluster sizes ($N_c \to \infty$). For finite $N_c$ (for the present study we use $N_c\!=\!4\!\times\!4$) it captures correlations up to the characteristic length scale of the cluster. 
For the D$\Gamma$A we employ its ladder-version in the particle-hole (magnetic) channel with Moriyaesque $\lambda$-corrections in the spin channel~\cite{ladderDGA,RohringerThesis,SchaeferThesis}. This choice of the scattering channel greatly simplifies the algorithm (as it bypasses the general, but complicated, parquet treatment) and is justified in the pseudogap regime of the Hubbard model, where fluctuation diagnostics methods could demonstrate unequivocally the dominance of the spin channel on the single-particle spectrum \cite{Gunnarsson2015,Wu2016, Rohringer2020,Schaefer2021b}.

Different from CDMFT, D$\Gamma$A captures short- and long-range fluctuations in the magnetic channel on equal footing which, as previous studies have shown, is indispensable in the vicinity of second order phase transitions \cite{Rohringer2011,Schaefer2017,Schaefer2019,Kitatani2019}. Moreover, D$\Gamma$A respects the Mermin-Wagner theorem \cite{Mermin1966, Hohenberg1967} and shows no ordering instability at finite temperatures for our two-dimensional model Eq.~(\ref{eq:Hubbard}). This is not the case for DMFT and CDMFT where the finite cluster size (for DMFT $N_c=1$) leads to an antiferromagnetic phase transition at a finite N{\' e}el temperature $T^{\text{N{\' e}el}}$. We therefore restrict ourselves to results obtained at temperatures above $T^{\text{N{\' e}el}}$ for these methods.

As impurity solver we use the latest generation of a continuous time quantum Monte-Carlo solver in its interaction expansion (CT-INT, \cite{Gull2011a}) which is an application of the TRIQS package \cite{TRIQS}.

\begin{figure*}[t!]
\begin{center}
\includegraphics[width=0.85\textwidth]{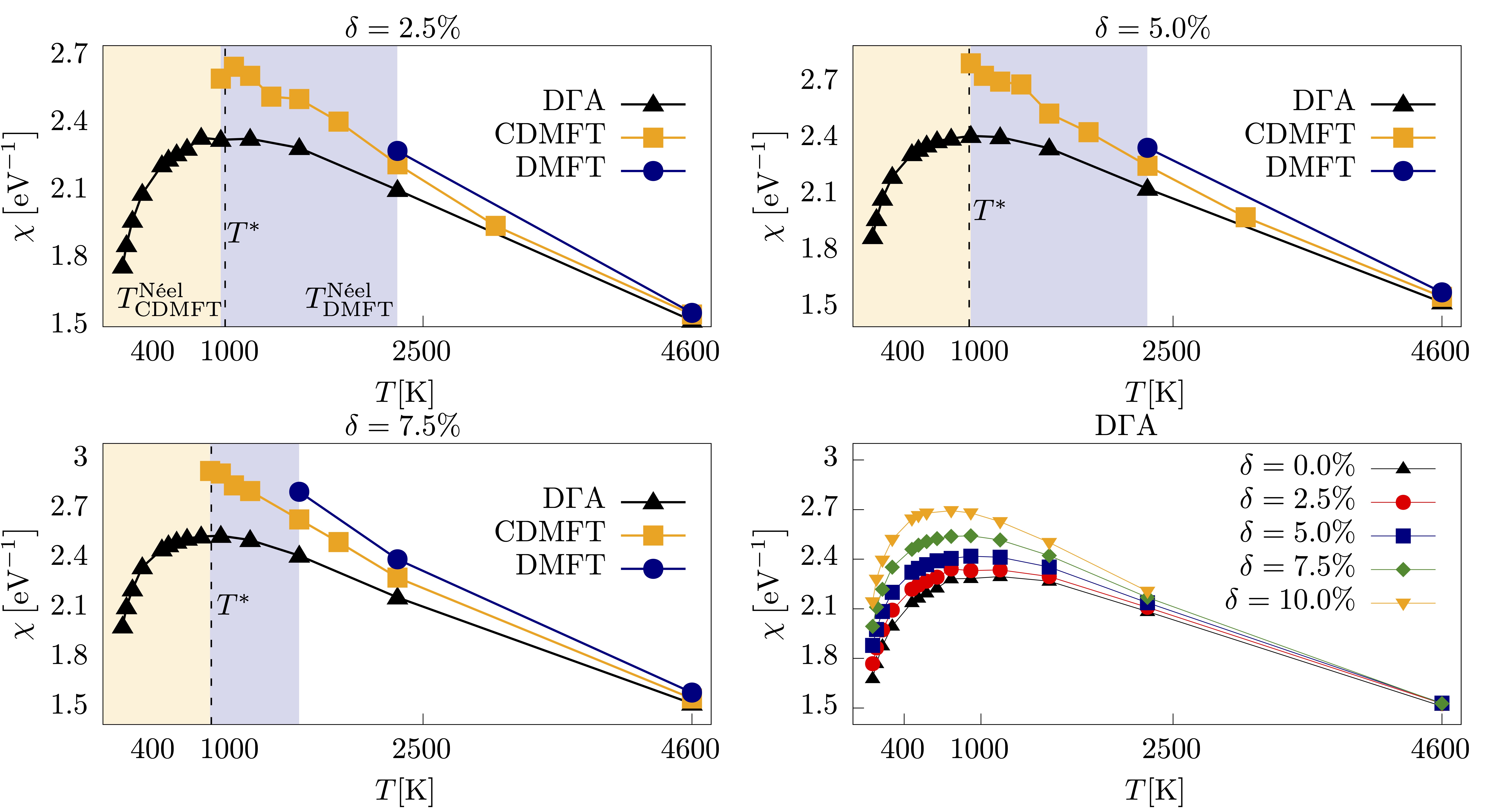}
\end{center}
\caption{Static uniform magnetic susceptibilities $\chi$ as a function of $T$ for 2.5\% (upper left panel), 5\% (upper right panel) and 7.5\% (lower left panel) hole doping. The transition temperature of DMFT $T_\textrm{DMFT}^{\text{N{\'e}el}}$ is indicated in blue, for CDMFT $T_\textrm{CDMFT}^{\text{N{\'e}el}}$ in orange. The lower right panel summarizes $\chi$ from D$\Gamma$A for different hole dopings.}\label{fig:2}
\end{figure*}

\section{Results}
\label{sec:results}
\subsection{Phase diagram and uniform susceptibilities}
\label{sec:phase_diagram}

We start the presentation of our results by discussing the phase diagram of Fig.~\ref{fig:1}, a summary of the data obtained by our different numerical techniques applied to Eq.~(\ref{eq:Hubbard}) as a function of doping ($\delta$ as the bottom horizontal axis, Sr-doping as the top one). In the left panel the black triangles represent the temperatures $T^*$ where the static uniform magnetic susceptibility $\chi\!\coloneqq\!\text{Re }\chi_{\text{m}}(\mathbf{q}\!=\!(0,0), i\Omega_n\!=\!0)$ displays a maximum in D$\Gamma$A. This temperature scale $T^*$ is highest in the half-filled case and monotonously decreases with increasing doping. Interestingly, for the doped system, this line follows to very good agreement the magnetic ordering temperature of CDMFT $T^{\text{N{\' e}el}}_\text{CDMFT}$, indicating the increased importance of non-local correlations. In the right panel we show the magnetic correlation length $\xi$ (red triangles) calculated with D$\Gamma$A for varying doping levels at the respective temperature $T^*$. $\xi(T^*)$ varies from around 1.2 to 2.5 lattice spacings (see also Sec.~\ref{sec:xi}).

For the determination of $T^*$ we turn to Fig.~\ref{fig:2}, which shows $\chi$ calculated by DMFT (red circles), CDMFT (orange squares) and D$\Gamma$A (black triangles) for three representative levels of doping. 
The shaded areas indicate magnetically ordered phases of DMFT (below $T_\text{DMFT}^\text{N{\'e}el}$) and CDMFT (below $T_\text{CDMFT}^\text{N{\'e}el}\!<\!T_\text{DMFT}^\text{N{\'e}el}$), respectively. In contrast, as $T_\text{D$\Gamma$A}^\text{N{\'e}el}=0$ we can trace $\chi$ obtained by D$\Gamma$A down to the lowest temperatures allowed by the impurity solver. Here, we determine its maximum at $T^*_\text{D$\Gamma$A}$ (shown as a black dashed line) by a third order polynomial fit of the numerical data. 
Overall we see that $\chi_\text{DMFT}\!>\!\chi_\text{CDMFT}\!>\!\chi_\text{D$\Gamma$A}$ which can be attributed to the increasing consideration of longer-ranged correlations in the approximation.
Next we observe that hole-doping away from half-filling reduces $T_\text{DMFT}^\text{N{\'e}el}$  \cite{Schaefer2017} and $T_\text{CDMFT}^\text{N{\'e}el}$ \cite{Fratino2017, Musshoff2021} (for the cluster size dependence of $T_\text{CDMFT}^\text{N{\'e}el}$ see \cite{Klett2020}).
In D$\Gamma$A, instead, the doping leads to a reduction of $T^*_{D\Gamma A}$ as highlighted in the bottom right panel of Fig.~\ref{fig:2}.

For all hole dopings considered in our nickelate model the flat maximum $\chi_\text{max}\!\coloneqq\!\chi(T^*)$ is a clear indicator of non-Curie-Weiss (and non-Pauli) behavior \cite{Huscroft01, Macridin06, Musshoff2021, Wietek2021b, Ortiz2021}. 
In high-T$_c$ cuprates such behavior is also seen in the suppression of the nuclear magnetic resonance (NMR) Knight shift \cite{Chen17}, which is the original hallmark of the onset of the pseudogap phase \cite{Alloul89}. Its second hallmark, observed in angle-resolved photoemission spectroscopy (ARPES, \cite{Damascelli2003}) is the non-isotropic suppression of spectral weight and emergence of Fermi arcs in the one-particle spectrum, which we investigate in the next section.

\subsection{Spectral functions}
\label{sec:spectral}
\begin{figure*}[t!]
\begin{center}
\includegraphics[width=0.85\textwidth]{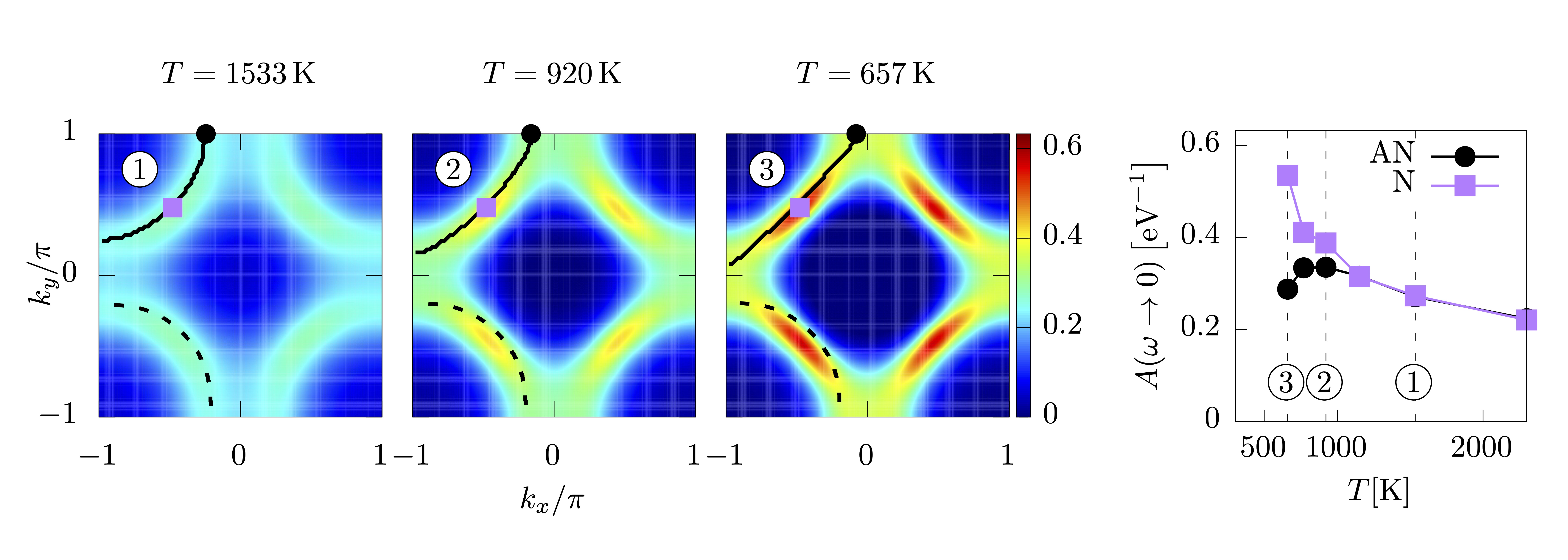}
\end{center}
\caption{\circle{1}--\circle{3}: Spectral intensities $A(\mathbf{k},\omega\!=\!0)$ for a constant doping of 7.5\% and temperatures of \circle{1} $1533\mathrm{K}$, \circle{2} $920 \mathrm{K}$, \circle{3}$ 657 \mathrm{K}$, calculated by D$\Gamma$A. The black lines indicate the Fermi surfaces of the non-interacting case (dashed) and interacting case (solid) [for increased readability, these are only shown for one quadrant of the Brillouin zone]. The nodal (purple square)-antinodal (black circle) differentiation of the spectral weight together with the suppression of it at the antinode (right-hand panel) is a clear indication of a pseudogap.}
\label{fig:3}
\end{figure*}

For the analysis of the one-particle spectral function $A(\mathbf{k},\omega\!=\!0)\!=\!-\frac{1}{\pi}\text{Im }G(\mathbf{k},i\omega_n\!\rightarrow\!0)$ in the paramagnetic phase we  restrict ourselves to a D$\Gamma$A analysis. 
Fig.~\ref{fig:3} shows the temperature evolution of $A(\mathbf{k},\omega\!=\!0)$ obtained by a linear fit of the first two Matsubara frequencies of the lattice Green function and extrapolation to zero frequency. Starting at the high-temperature point \circle{1} at $T\!=\!1533$ K we follow the \emph{vertical} dashed line (i.e. at fixed 7.5\% hole doping) in the $T$/$\delta$ phase diagram of Fig.~\ref{fig:1}. At the highest temperature \circle{1} both the spectral intensity (indicated by the color scale) as well as the interacting Fermi surface (solid black line) follow the (hole-like) shape of the non-interacting Fermi surface (dashed black line). The locations of the Fermi surface points have been obtained from the roots of the quasi-particle equation (QPE)
\begin{equation}
    \tilde{\varepsilon}(\mathbf{k})\!\coloneqq\!\varepsilon(\mathbf{k}) - \mu + \textrm{Re }\Sigma(\mathbf{k}, \mathrm{i} \omega_n\!\to\!0),
    \label{eq:qpe}
\end{equation}
where $\varepsilon(\mathbf{k})$ is the non-interacting dispersion relation and $\Sigma(\mathbf{k},\mathrm{i\omega_n})$ the self-energy from D$\Gamma$A (which is zero in the non-interacting case).
Cooling the system across $T^*$, and passing \circle{2} $920$ K and \circle{3} $657$ K, one first notices that the shape of the Fermi surface starts to deviate strongly from the non-interacting case. This can be attributed to self-energy effects stemming from non-local correlations. This behaviour is also in qualitative agreement with recent numerically exact diagrammatic Monte Carlo calculations \cite{Rossi2020} for smaller interactions. 
Second, at low temperatures, one can observe a clear Fermi arc structure of the spectral intensity. 
Third, the temperature dependence of the spectral weight at the antinode (black circle) starts to differ strongly from that of the node (purple square): At high temperatures, both values increase when the system is cooled. After reaching $T^*$, however, only the spectral weight at the node continues to grow, whereas at the antinode it starts to decrease. 
Together with the decrease in the uniform static magnetic susceptibility (see Sec.~\ref{sec:phase_diagram}) this is an unequivocal indication of the onset of a pseudogap regime. Topologically at $T^*$ the Fermi surface is hole-like, i.e. $\tilde{\varepsilon}(\mathbf{k}\!=\!(\pi,0))\!<\!0$ in Eq.~(\ref{eq:qpe}), in accordance with the finding of \cite{Wu2018} that a pseudogap develops only for hole-like Fermi surface topologies. 
We also note that our results for $T^*$ agree for small dopings with the ones obtained within the dynamical cluster approximation (DCA) on eight sites with similar model parameters \cite{Wu2018}, however, the drop with doping is less pronounced within the compared doping range for our data. We sense that this is an effect of both, slightly different model parameters and the DCA momentum patching, which for this model and small cluster sizes is not able to resolve the exact location of the antinode away from $\mathbf{k}\!=\!(\pi,0)$. This resolution, however, is possible within D$\Gamma$A so that the location of antinode and node can be precisely determined within the Brillouin zone [e.g. $\mathbf{k}_\text{AN}\!=\!(\pi,0.51)$ and $\mathbf{k}_\text{N}\!=\!(1.51,1.51)$ for \circle{2}].

\begin{figure*}[t!]
\begin{center}
\includegraphics[width=0.85\textwidth]{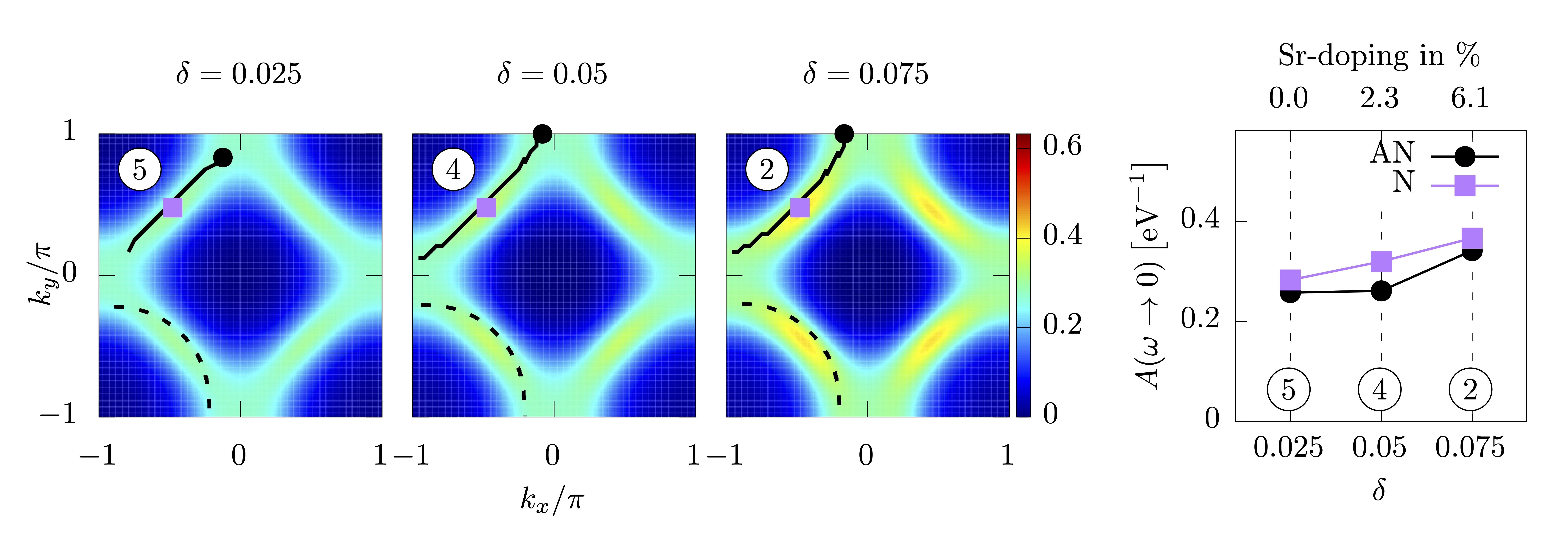}
\end{center}
\caption{Analogous plots to Fig.~\ref{fig:3} at fixed $T\!=\!920$ K for dopings \circle{5} 2.5\%, \circle{4} 5\% and \circle{2} 7.5\%.}
\label{fig:4}
\end{figure*}

In Fig.~\ref{fig:4} we show the complementary evolution of the spectral intensity across the $T^*$ line at fixed $T\!=\!920$ K following the \emph{horizontal} dashed line in phase diagram Fig.~\ref{fig:1}. 
On can observe here that the progressive reduction of the doping from \circle{2} 7.5\% to \circle{5} 2.5\% leads to a significant drop of spectral intensity at both the node and the antinode and, eventually, to a mitigation of the nodal-antinodal differentiation. Furthermore, there is a strong tendency visible toward a reconstruction of the topology of the Fermi surface from hole-like [$\tilde{\varepsilon}(\mathbf{k}\!=\!(\pi,0))\!<\!0$, within the pseudogap regime] to electron-like [$\tilde{\varepsilon}(\mathbf{k}\!=\!(\pi,0))\!>\!0$, at small dopings \cite{Wu2018,Scheurer2018}], what can be attributed to strong non-local fluctuations approaching half-filling.

In order to investigate more closely the nature of the emerging pseudogap, in the next section we analyse the magnetic correlation length and the  momentum-dependent magnetic response.

\subsection{Momentum-dependent susceptibility and correlation lengths}
\label{sec:xi}

\begin{figure*}[t!]
\begin{center}
\includegraphics[width=0.85\textwidth]{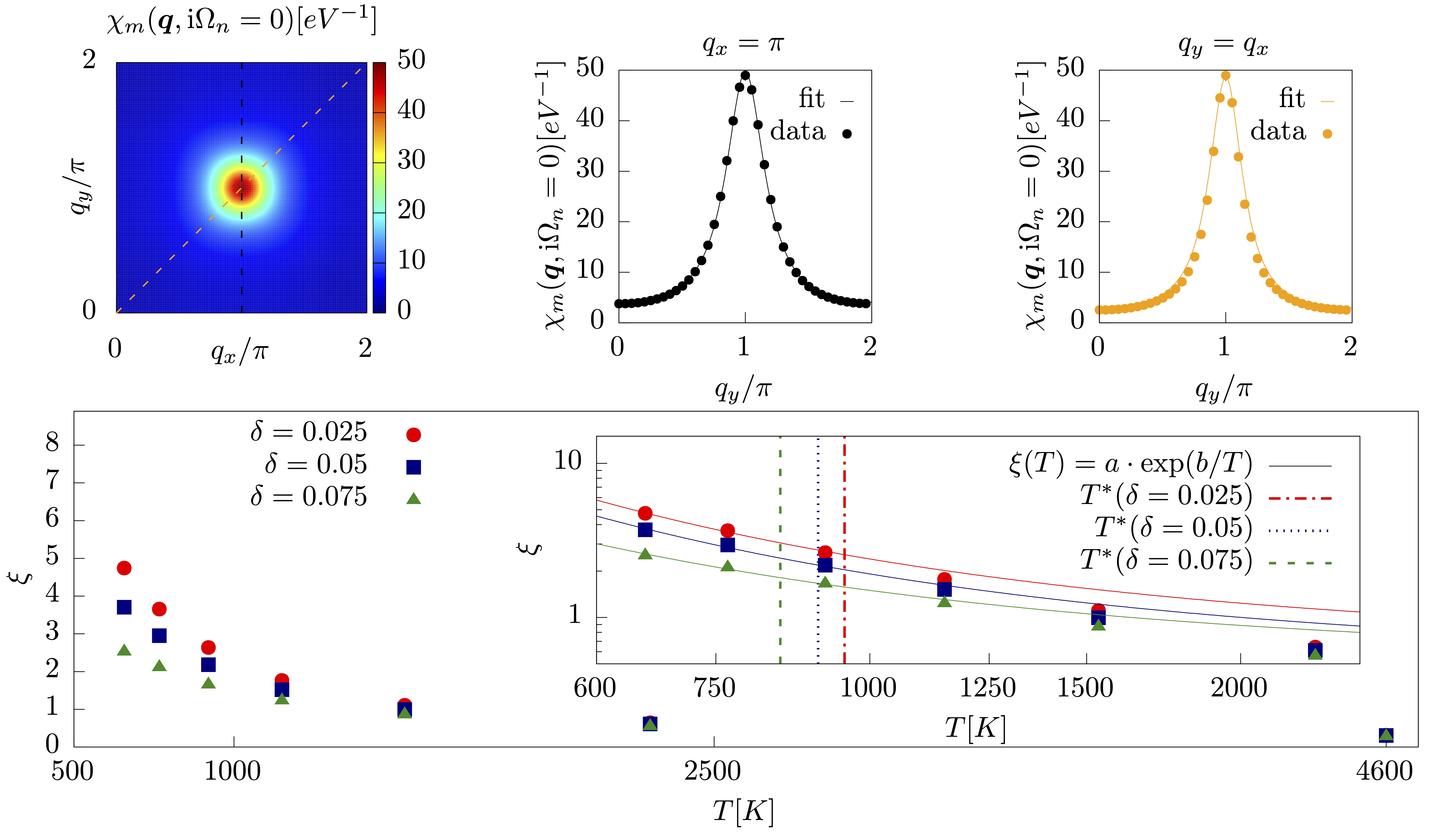}
\end{center}
\caption{Top, from left to right: $\chi_{\text{m}}(\mathbf{q},i\Omega_n\!=\!0)$, $\chi_{\text{m}}(\mathbf{q}\!=\!(q_x,\pi),i\Omega_n\!=\!0)$ and $\chi_{\text{m}}(\mathbf{q}\!=\!(q_x,q_x),i\Omega_n\!=\!0)$ for a doping of $\delta\!=\!7.5\%$ and a temperature of $960 \mathrm{K}$, calculated by D$\Gamma$A. Circles denote calculated points, the solid line an Ornstein-Zernike fit by Eq.~(\ref{eqn:ornstein_sin}). Bottom: Correlation lengths $\xi$ of D$\Gamma$A plotted over the temperature for different dopings. The insets shows a double-logarithmic plot and temperature fits [see text and Eq.~(\ref{eqn:xi})].}
\label{fig:5}
\end{figure*}

We first calculate the fully momentum-dependent static magnetic susceptibility $\chi_{\text{m}}(\mathbf{q},i\Omega_n\!=\!0)$ within D$\Gamma$A. The top leftmost panel of Fig.~\ref{fig:5} shows results for \circle{2} ($\delta\!=\!7.5\%$ and $T\!=\!960$ K), which is slightly above $T^*$ for this doping. The maximum value of $\chi_{\text{m}}(\mathbf{q},i\Omega_n\!=\!0)$ is assumed at $\mathbf{q}\!=\!\mathbf{Q}\!=\!(\pi,\pi)$ at $T^*$. We note in passing that also incommensurate N{\'e}el order with $\mathbf{Q}\!\neq\!(\pi,\pi)$ may occur in different parameter regimes of the model  \cite{Schaefer2017,Huang2018,Wietek2021b,Simkovic2021}. For obtaining the correlation length $\xi$ we perform an Ornstein-Zernike fit with \cite{Zernike1916,Rohringer2011,Schaefer2015,Schaefer2021}
\begin{eqnarray*}
 \chi_{\text{m}}(\mathbf{q}, i\Omega_{n}=0)&=&\frac{A}{4\text{sin}^{2}\left(\frac{q_x-Q_x}{2}\right) +  4\text{sin}^{2}\left(\frac{q_y-Q_y}{2}\right) + \xi^{-2}}\\&\overset{\mathbf{q}\rightarrow{\mathbf{Q}}}{\rightarrow}&\frac{A}{(\mathbf{q}-\mathbf{Q})^2+\xi^{-2}},
 \label{eqn:ornstein_sin}
\end{eqnarray*}
where $\mathbf{Q}$ denotes the momentum vector where the susceptibility assumes its maximum value. Assuming this functional form for the fit is justified by two exemplary fits in the momentum directions $\mathbf{q}\!=\!(q_x,\pi)$ and $\mathbf{q}\!=\!(q_x,q_x)$ shown in the upper center and right panels of Fig.~\ref{fig:5}.   
The so-obtained temperature dependence of $\xi$ for several dopings is plotted in the lower panel of Fig.~\ref{fig:5}. For small dopings we fit this dependence with
\begin{equation}
    \xi\!=\!\xi_0 e^{2\pi\rho_S/T}
    \label{eqn:xi}
\end{equation}
(with $\rho_S$ being the spin stiffness), characteristic of a low-$T$ gapped regime in two dimensions. The fit works reasonably well for temperatures $T\!<\!T^*$, hinting towards a magnetically ordered ground state in D$\Gamma$A for the dopings investigated.

As already commented in the discussion of Fig.~\ref{fig:1}, the correlation lengths at the pseudogap temperature $T^*$ range from 1.2 to about 2 lattice spacings. This is a clear indicator that the pseudogap mechanism in our case is not the one observed in the weak coupling regime of the Hubbard model \cite{Vilk1996,Vilk1997,Schaefer2015,Schaefer2016,Simkovic2020,Kim2020,Hille2020,Schaefer2021}: there, in contrast, the pseudogap is opened when the magnetic correlation length exceeds the the thermal de Broglie wavelength of the quasiparticles $\xi \gg v_\text{F}/(\pi T)$ (Vilk criterion), where $v_\text{F}$ is the Fermi velocity. Hence, in the weak coupling regime, large correlation lengths have to be present for opening the (pseudo-)gap. This, however, does not need to be the case for stronger coupling: here, already the treatment of short-ranged (spin) fluctuations allows for the development of a pseudogap as momentum-differentiated gap, which is the reason for the successful description of this regime by cluster extensions of DMFT (like CDMFT and DCA \cite{Huscroft01,Macridin06,Kyung2006,Gull2013,Gunnarsson2015,Fratino2021}).

\section{Discussion and conclusions}
\label{sec:conclusions}
To summarize, we analyzed a material-realistic single-band Hubbard model for the infinite-layer nickelate compound LaNiO$_2$.
By a combination of cellular dynamical mean-field theory and dynamical vertex approximation calculations we could trace the temperatures sufficiently low to determine a flat maximum in the uniform static magnetic susceptibility for the hole-doped system at $T^*$. 
This temperature marks the onset of the pseudogap regime which manifests on the one-particle level as Fermi arcs in the spectral function. Concomitant on the two-particle level, the momentum-resolved magnetic susceptibility shows short-ranged magnetic fluctuations, which is characteristic of a strong coupling pseudogap. 
The exact location of the change from a 
weak-coupling to a strong-coupling pseudogap regime is a matter of current debate. Three indicators for this change can be mentioned: (i) a sudden increase in electronic correlations leading to a change in Fermi surface topology \cite{Wu2018}, (ii) this strong correlation regime hosts relatively short-ranged correlations with the occurrence of (partial) localization \cite{Simkovic2021,Gunnarsson2015} and (iii) the electron-boson coupling vertex develops a significant imaginary part \cite{Krien2021,vanLoon2018}. 
Our investigations of (i) and (ii) in this manuscript by means of the D$\Gamma$A, hence, allow us to characterize the found pseudogap as driven by strong coupling (Mott) physics.

In conclusion, our results for LaNiO$_2$ support the idea that the infinite-layer nickelates and new nickelate superconductors are indeed close relatives of other unconventional superconductors and, in particular, high-T$_\text{c}$ cuprates. This is a most promising perspective as contrasting nickelates with cuprates %superconductors 
might lead to a much deeper understanding of non-phonon mediated pairing. 
Indeed future research should focus on apparent differences between the two material classes. Specifically, the absence of magnetic order in the infinite-layer nickelate compounds as well as their reduced covalency with oxygen \cite{lee04, Hepting2020} compared to the cuprates is remarkable. 
Whether this means that also pairing mechanisms are distinct remains to be investigated.

\section*{Conflict of Interest Statement}
The authors declare that the research was conducted in the absence of any commercial or financial relationships that could be construed as a potential conflict of interest.

\section*{Author Contributions}
M.~Klett performed the numerical calculations and the post-processing of the data. The manuscript has been written by M.~Klett, P.~Hansmann, and  T.~Sch{\"a}fer. T.~Sch{\"a}fer initiated and supervised the project. All authors provided critical feedback and shaped the research, analysis and manuscript.

\section*{Acknowledgments}
We thank K.~Held, M.~Kitatani, L.~Si, P.~Worm, M.~Hepting, M.~Ferrero, A.~Georges, F.~{\v S}imkovic and A.~Toschi for insightful discussions and F.~{\v S}imkovic and E.~K{\" o}nig for critically reading the manuscript. We thank the computing service facility of the MPI-FKF for their support and we gratefully acknowledge use of the computational resources of the Max Planck Computing and Data Facility.

\section*{Computational details}
For applying D$\Gamma$A with Moriyaesque $\lambda$-corrections we solve the Bethe-Salpeter equations in Matsubara frequency space with $N_{i\omega}\!=\!90$ positive fermionic and $N_{i\Omega}\!=\!89$ positive bosonic Matsubara frequencies for the two-particle Green function at all temperature shown, as well as $200$ linear momentum grid points. To converge the DMFT calculation self-consistently we used the continuous-time quantum Monte Carlo solver in an interaction expansion (CT-INT) as part of an application of the TRIQS package. Every iteration was done using $256 \cdot 10^5$ cycles and roughly $6200$ core hours per temperature. For all shown spectral function plots over the Brillouin Zone we used a momentum resolution of $3000$ k-points.

In order to extract the magnetic susceptibility in a CDMFT approach we apply a ferromagnetic field on each lattice site with field strengths $H_F = 0.02,0.04,0.06$ to get the slope of a linear fit which enables us to calculate the magnetic susceptibility:
\begin{equation}
    \text{Re }\chi_m(\mathbf{q}\!=\!(0,0),i\Omega_n\!=\!0) = \left.\frac{\partial m}{\partial H}\right\vert_{H=0} \approx \frac{m}{H_F}.
\end{equation}
For each doping/temperature point we checked that all the applied fields are still within the linear response regime. All CDMFT calculations are again performed using the CT-INT quantum Monte Carlo solver in a continuous-time approach. For every data point we used twenty self-consistency steps, each using $6.4$ million Monte Carlo cycles. 

\bibliographystyle{apsrev4-2}
\bibliography{references}

\end{document}